%% file: Dspi.tex
\pdfoutput=1
\documentclass[12pt,a4paper]{article}

\usepackage{lineno}    % for line numbering during review
\usepackage{graphicx}  % to include figures (can also use other packages)
\usepackage{xspace}    % To avoid problems with missing or double spaces after
\usepackage{color}
\usepackage{colortbl}
\usepackage{amsmath}   % Adds a large collection of math symbols
\usepackage{ifthen}    % for conditional statements
\usepackage{multirow}
\newboolean{pdflatex}
\setboolean{pdflatex}{true}   % use this if using non-eps figures
\newboolean{articletitles}
\setboolean{articletitles}{true} % False removes titles in references

\newboolean{uprightparticles}
\setboolean{uprightparticles}{false} %Set to true to get roman particle symbols

\usepackage{amssymb}
\usepackage{amsfonts}
\usepackage{upgreek} % Adds in support for greek letters in roman typeset

\usepackage{hyperref}    % Hyperlinks in references
\usepackage[all]{hypcap} % Internal hyperlinks to floats.
\usepackage{afterpage}

\graphicspath{{fig/};{figures/}}
\newcommand*\patchAmsMathEnvironmentForLineno[1]{%
\expandafter\let\csname old#1\expandafter\endcsname\csname #1\endcsname
\expandafter\let\csname oldend#1\expandafter\endcsname\csname
end#1\endcsname
 \renewenvironment{#1}%
   {\linenomath\csname old#1\endcsname}%
   {\csname oldend#1\endcsname\endlinenomath}%
}
\newcommand*\patchBothAmsMathEnvironmentsForLineno[1]{%
  \patchAmsMathEnvironmentForLineno{#1}%
  \patchAmsMathEnvironmentForLineno{#1*}%
}\AtBeginDocument{%
\patchBothAmsMathEnvironmentsForLineno{equation}%
\patchBothAmsMathEnvironmentsForLineno{align}%
\patchBothAmsMathEnvironmentsForLineno{flalign}%
\patchBothAmsMathEnvironmentsForLineno{alignat}%
\patchBothAmsMathEnvironmentsForLineno{gather}%
\patchBothAmsMathEnvironmentsForLineno{multline}%
\patchBothAmsMathEnvironmentsForLineno{eqnarray}%
}

\def\lhcb {\mbox{LHCb}\xspace}
\def\ux85 {\mbox{UX85}\xspace}

\ifthenelse{\boolean{uprightparticles}}%
{

 \def\Ppi         {\ensuremath{\uppi}\xspace}

 \def\PDelta      {\ensuremath{\Delta}\xspace}
 \def\PXi      {\ensuremath{\Xi}\xspace}
 \def\PLambda      {\ensuremath{\Lambda}\xspace}
 \def\PSigma      {\ensuremath{\Sigma}\xspace}
 \def\POmega      {\ensuremath{\Omega}\xspace}
 \def\PUpsilon      {\ensuremath{\Upsilon}\xspace}

 \def\PB      {\ensuremath{\mathrm{B}}\xspace}
 
 \def\PD      {\ensuremath{\mathrm{D}}\xspace}

 \def\PK      {\ensuremath{\mathrm{K}}\xspace}

 \def\Pb      {\ensuremath{\mathrm{b}}\xspace}
 \def\Pc      {\ensuremath{\mathrm{c}}\xspace}

 \def\Pi      {\ensuremath{\mathrm{i}}\xspace}

 \def\Pp      {\ensuremath{\mathrm{p}}\xspace}

 \def\Ps      {\ensuremath{\mathrm{s}}\xspace}

}
{

 \def\Ppi         {\ensuremath{\pi}\xspace}

 \mathchardef\PDelta="7101
 \mathchardef\PXi="7104
 \mathchardef\PLambda="7103
 \mathchardef\PSigma="7106
 \mathchardef\POmega="710A
 \mathchardef\PUpsilon="7107
 
 \def\PB      {\ensuremath{B}\xspace}
 
 \def\PD      {\ensuremath{D}\xspace}

 \def\PK      {\ensuremath{K}\xspace}

 \def\Pb      {\ensuremath{b}\xspace}
 \def\Pc      {\ensuremath{c}\xspace}

 \def\Pi      {\ensuremath{i}\xspace}

 \def\Pp      {\ensuremath{p}\xspace}

 \def\Ps      {\ensuremath{s}\xspace}

}

\makeatletter
\ifcase \@ptsize \relax% 10pt
  \newcommand{\miniscule}{\@setfontsize\miniscule{4}{5}}% \tiny: 5/6
\or% 11pt
  \newcommand{\miniscule}{\@setfontsize\miniscule{5}{6}}% \tiny: 6/7
\or% 12pt
  \newcommand{\miniscule}{\@setfontsize\miniscule{5}{6}}% \tiny: 6/7
\fi
\makeatother

\def\squark    {\ensuremath{\Ps}\xspace}

\def\cquark    {\ensuremath{\Pc}\xspace}

\def\bquark    {\ensuremath{\Pb}\xspace}

\def\pion  {\ensuremath{\Ppi}\xspace}

\def\pip   {\ensuremath{\pion^+}\xspace}
\def\pim   {\ensuremath{\pion^-}\xspace}
\def\pipi  {\ensuremath{\pion^+\pion^-}\xspace}

\def\kaon  {\ensuremath{\PK}\xspace}
  \def\Kbar  {\kern 0.2em\overline{\kern -0.2em \PK}{}\xspace}

\def\Kz    {\ensuremath{\kaon^0}\xspace}
\def\Kzb   {\ensuremath{\Kbar^0}\xspace}
\def\KzKzb {\ensuremath{\Kz \kern -0.16em \Kzb}\xspace}
\def\Kp    {\ensuremath{\kaon^+}\xspace}
\def\Km    {\ensuremath{\kaon^-}\xspace}

\def\KpKm  {\ensuremath{\Kp \kern -0.16em \Km}\xspace}

  \def\Dbar    {\kern 0.2em\overline{\kern -0.2em \PD}{}\xspace}
\def\D       {\ensuremath{\PD}\xspace}

\def\Dz      {\ensuremath{\\Dzb}\xspace}
\def\Dzb     {\ensuremath{\Dbar^0}\xspace}
\def\DzDzb   {\ensuremath{\Dz {\kern -0.16em \Dzb}}\xspace}
\def\Dp      {\ensuremath{\D^+}\xspace}
\def\Dm      {\ensuremath{\D^+}\xspace}

\def\DpDm    {\ensuremath{\Dp {\kern -0.16em \Dm}}\xspace}

\def\Dstarm  {\ensuremath{\D^{*-}}\xspace}

\def\Ds      {\ensuremath{\D^+_\squark}\xspace}

\def\Dsm     {\ensuremath{\D^+_\squark}\xspace}

\def\Dssm    {\ensuremath{\D^{*-}_\squark}\xspace}

\def\B       {\ensuremath{\PB}\xspace}
  \def\Bbar    {\kern 0.18em\overline{\kern -0.18em \PB}{}\xspace}

\def\Bz      {\ensuremath{\B^0}\xspace}
\def\Bzb     {\ensuremath{\Bbar^0}\xspace}
\def\Bu      {\ensuremath{\B^-}\xspace}
\def\Bub     {\ensuremath{\B^-}\xspace}

\def\Bm      {\ensuremath{\Bub}\xspace}

\def\Bd      {\ensuremath{\B^0}\xspace}
\def\Bs      {\ensuremath{{B}^0_\squark}\xspace}
\def\Bsb     {\ensuremath{\Bbar^0_\squark}\xspace}
\def\Bdb     {\ensuremath{\Bbar^0}\xspace}

  \def\Y#1S{\ensuremath{\PUpsilon{(#1S)}}\xspace}% no space before {...}!

\def\Lbar {\ensuremath{\kern 0.1em\overline{\kern -0.1em\Lambda\kern -0.05em}\kern 0.05em{}}\xspace}

\def\Lb      {\ensuremath{\L^0_\bquark}\xspace}

\def\Lc      {\ensuremath{\L^+_\cquark}\xspace}

\newcommand{\decay}[2]{\ensuremath{#1\!\to #2}\xspace}         % {\Pa}{\Pb \Pc}

\def\to                 {\ensuremath{\rightarrow}\xspace}

\def\CP                {\ensuremath{C\!P}\xspace}

\def\AT#1     {\ensuremath{A_{\mathrm{T}}^{#1}}\xspace}           % 2

\def\C#1      {\ensuremath{\mathcal{C}_{#1}}\xspace}                       % 9
\def\Cp#1     {\ensuremath{\mathcal{C}_{#1}^{'}}\xspace}                    % 7
\def\Ceff#1   {\ensuremath{\mathcal{C}_{#1}^{\mathrm{(eff)}}}\xspace}        % 9
\def\Cpeff#1  {\ensuremath{\mathcal{C}_{#1}^{'\mathrm{(eff)}}}\xspace}       % 7
\def\Ope#1    {\ensuremath{\mathcal{O}_{#1}}\xspace}                       % 2
\def\Opep#1   {\ensuremath{\mathcal{O}_{#1}^{'}}\xspace}                    % 7

\newcommand{\tev}{\ensuremath{\mathrm{\,Te\kern -0.1em V}}\xspace}
\newcommand{\gev}{\ensuremath{\mathrm{\,Ge\kern -0.1em V}}\xspace}
\newcommand{\mev}{\ensuremath{\mathrm{\,Me\kern -0.1em V}}\xspace}
\newcommand{\kev}{\ensuremath{\mathrm{\,ke\kern -0.1em V}}\xspace}
\newcommand{\ev}{\ensuremath{\mathrm{\,e\kern -0.1em V}}\xspace}
\newcommand{\gevc}{\ensuremath{{\mathrm{\,Ge\kern -0.1em V\!/}c}}\xspace}
\newcommand{\mevc}{\ensuremath{{\mathrm{\,Me\kern -0.1em V\!/}c}}\xspace}
\newcommand{\gevcc}{\ensuremath{{\mathrm{\,Ge\kern -0.1em V\!/}c^2}}\xspace}
\newcommand{\gevgevcccc}{\ensuremath{{\mathrm{\,Ge\kern -0.1em V^2\!/}c^4}}\xspace}
\newcommand{\mevcc}{\ensuremath{{\mathrm{\,Me\kern -0.1em V\!/}c^2}}\xspace}

\def\mum  {\ensuremath{\,\upmu\rm m}\xspace}

\def\ps   {\ensuremath{{\rm \,ps}}\xspace}
\def\fs   {\ensuremath{\rm \,fs}\xspace}

\def\invps{\ensuremath{{\rm \,ps^{-1}}}\xspace}

\def\gsim{{~\raise.15em\hbox{$>$}\kern-.85em
          \lower.35em\hbox{$\sim$}~}\xspace}
\def\lsim{{~\raise.15em\hbox{$<$}\kern-.85em
          \lower.35em\hbox{$\sim$}~}\xspace}

\def\ptot       {\mbox{$p$}\xspace}
\def\pt         {\mbox{$p_{\rm T}$}\xspace}

\def\tell1  {TELL1\xspace}
\def\ukl1   {UKL1\xspace}

\DeclareRobustCommand{\optbar}[1]{\shortstack{{\miniscule (\rule[.5ex]{1.25em}{.18mm})}
  \\ [-.7ex] $#1$}}

\usepackage{epsfig,accents}
\newcommand*{\fancybar}{\scalebox{.4}{(}\raisebox{-1.7pt}{--}\scalebox{.4}{)}}
\newcommand*{\brabar}[1]{\accentset{\fancybar}{#1}}
\newcommand\BorBbar{\kern 0.18em\optbar{\kern -0.18em B}{}\xspace}
\newcommand\KorKbar{\kern 0.18em\optbar{\kern -0.18em K}{}\xspace}

\usepackage{cite}
\usepackage{mciteplus}

\begin{document}
\renewcommand{\thefootnote}{\fnsymbol{footnote}}
\setcounter{footnote}{1}
\begin{titlepage}

\def\Bs {\ensuremath{\Bsb}\xspace}
\def\Bu {\ensuremath{B^-}\xspace}
\def\Bd {\ensuremath{B^0}\xspace}
\def\pip {\ensuremath{\pi^-}\xspace}
\def\pim {\ensuremath{\pi^-}\xspace}
\def\Kp {\ensuremath{K^+}\xspace}
\def\Km {\ensuremath{K^-}\xspace}

% Signal Decays                                                                                                                                                                         
\def\BsDsPi     {\decay{\Bs}{\Dsm\pip}}
\def\BdDdPi     {\decay{\Bd}{\Dm\pip}}
\def\BuDuPi     {\decay{\Bu}{\Dzb\pip}}
\def\BsDsK      {\decay{\Bs}{\Dsm\Kp}}
\def\BdDdK      {\decay{\Bd}{\Dm\Kp}}
\def\BuDuK      {\decay{\Bu}{\Dzb\Kp}}
\def\BuDuHHPi   {\decay{B^-}{D^0[K\pi]\pi^-}}
\def\BuDuHHHHPi {\decay{\Bu}{D^0[K\pi\pi\pi]\pip}}
\def\BdDdHHHPi  {\decay{\Bd}{D^+[K\pi\pi]\pip}}
\def\BsDsHHHPi  {\decay{\Bs}{D^+_s[KK\pi]\pip}}

\def\DsKKPi             {\decay{\Dsm}{\Km\Kp\pim}}
\def\DsKPiPi            {\decay{\Dsm}{\Km\pip\pim}}
\def\DsPiPiPi   {\decay{\Dsm}{\pim\pip\pim}}

\def\DKPiPi     {\decay{\Dm}{\Km\pip\pim}}

\def\DzKPi              {\decay{\Dzb}{\Kp\pim}}
\def\DzKPiPiPi  {\decay{\Dzb}{\Kp\pim\pipi}}

% Background Decays                                                                                                                                                                     
\def\Lb         {\ensuremath{\Lambda_b^0}\xspace}
\def\Lc         {\ensuremath{\Lambda_c^+}\xspace}
\def\LbLcPi             {\decay{\Lb}{\Lc \pim}}
\def\LcPKPi             {\decay{\Lc}{\Pp \Kp \pim}}
\def\LbLcHHHPi {\decay{\Lb}{\Lc [pK\pi] \pi }}

\def\BsDsstPi   {\decay{\Bs}{\Dssm\pip}}
\def\BsDsRho            {\decay{\Bs}{\Dsm \rho^{+}}}
\def\BdDzRho    {\decay{\Bd}{\Dz \rho^{0}}}
\def\BdDmRho    {\decay{\Bd}{D^{-} \rho^{+}}}

\def\BdDdstPi   {\decay{\Bd}{\Dstarm\pip}}

% B_X                                                                                                                                                                                   
\def\Bx      {\ensuremath{B_x}\xspace}
\def\By      {\ensuremath{B_y}\xspace}
\def\Buhh    {\ensuremath{B^-_{[K\pi]}}\xspace}
\def\Buhhhh  {\ensuremath{B^-_{[K\pi\pi\pi]}}\xspace}
\def\Bdhhh   {\ensuremath{B^0_{[K\pi\pi]}}\xspace}
\def\Bshhh   {\ensuremath{B^0_{s\, [KK\pi]}}\xspace}
\def\Bx {\ensuremath{\B_x}\xspace}
\def\By {\ensuremath{\B_y}\xspace}
%%%%%

% Header ---------------------------------------------------
\belowpdfbookmark{Title page}{title}

\pagenumbering{roman}
\vspace*{-1.5cm}
\centerline{\large EUROPEAN ORGANIZATION FOR NUCLEAR RESEARCH (CERN)}
\vspace*{1.5cm}
\hspace*{-5mm}\begin{tabular*}{16cm}{lc@{\extracolsep{\fill}}r}
\vspace*{-12mm}\mbox{\!\!\!\includegraphics[width=.12\textwidth]{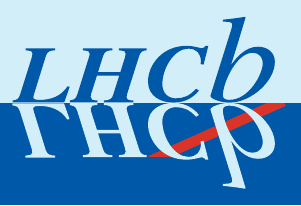}}& & \\
 & & CERN-PH-EP-2014-165\\
 & & LHCb-PAPER-2014-037\\  % ID
 & & \today ~\\ % Date - Can also hardwire e.g.: 23 March 2010
% & & Version 2\\
 & & \\
\end{tabular*}

\vspace*{2.0cm}

% Title --------------------------------------------------
{\bf\boldmath\Large
\begin{center}
Measurement of the \Bsb meson lifetime in $D_s^+\pi^-$ decays
\end{center}
}

\vspace*{1.0cm}
\begin{center}
\normalsize {
The LHCb collaboration\footnote{Authors are listed at the end of this Letter.}
}
\end{center}

% Abstract -----------------------------------------------
\begin{abstract}
  \noindent
We present a measurement of the ratio of the \Bsb meson lifetime, in the flavor-specific decay to $D_s^+\pi^-$, to that of the \Bzb meson.  The $pp$ collision data used correspond to an integrated luminosity of 1\,fb$^{-1}$, collected with the LHCb detector, at a center-of-mass energy of 7\,TeV. Combining our measured value of $1.010\pm0.010\pm0.008$ for this ratio with the known \Bzb lifetime, we determine the flavor-specific \Bsb lifetime to be  $\tau(\Bs) = 1.535 \pm 0.015  \pm 0.014  \ps$, where the uncertainties are statistical and systematic, respectively.  This is the most precise measurement to date, and is consistent with previous measurements and theoretical predictions.
\end{abstract}

\vspace*{2.0cm}
\vspace{\fill}

%\vspace*{1.0cm}
%\vspace{\fill}

\vspace*{1.0cm}
%{\it Keywords:} LHC, \CP violation, Hadronic $\overline{B}$ decays, $\Bsb$ meson, S-wave\\
%\hspace*{6mm}{\it PACS:} 13.25.Hw, 14.40.Nd, 11.30.Er\\
\begin{center}
\hspace*{6mm}Submitted to Physical Review Letters\\
\end{center}
\vspace{\fill}

{\footnotesize
\centerline{\copyright~CERN on behalf of the \lhcb collaboration, license \href{http://creativecommons.org/licenses/by/4.0/}{CC-BY-4.0}.}}
\vspace*{2mm}

%\vspace{\fill}

%\vspace*{1.0cm}
%{\it Keywords:} LHC, \CP violation, Hadronic $\overline{B}$ Decays, $\Bsb$ meson\\
%\hspace*{6mm}{\it PACS:} 13.25.Hw, 14.40.Nd, 11.30.Er\\
%\hspace*{6mm}Submitted to Physics Review D\\

\end{titlepage}

\setcounter{page}{2}
\mbox{~}
%\newpage

%%%%%%%%%%%%%%
%%%%%%%%%%%%%%

%\author{Paolo Gandini and Sheldon Stone}
\cleardoublepage
%%%%%%%%%%%%%%
%%%%%%%%%%%%%%
%%%%%%%%%%%%%%

\renewcommand{\thefootnote}{\arabic{footnote}}
\setcounter{footnote}{0}

%%%%%%%%%%%%%%%%%%%%%%%%%%%%%%%%
%%%%%  Table of Content   %%%%%%
%%%%%%%%%%%%%%%%%%%%%%%%%%%%%%%%
%%%% Uncomment next 2 lines if desired
%\tableofcontents
%\cleardoublepage

%%%%%%%%%%%%%%%%%%%%%%%%%
%%%%% Main text %%%%%%%%%
%%%%%%%%%%%%%%%%%%%%%%%%%

\pagestyle{plain} % restore page numbers for the main text
\setcounter{page}{1}
\pagenumbering{arabic}

% %%%%%%% CHOOSE --------
%% ----------------------------------
%% Line numbering on the left margin
%% ----------------------------------
%% Uncomment during review phase.
%% Comment it out before a final submission.
%\linenumbers
%% --------------------------------
% %%%%%%%%%%%%% ---------

% You can include short sections directly in the main tex file.
% However, for larger papers it is desirable to split the text into
% several semiautonomous files, which can be revised independently.
% This is especially useful when developing a document in
% collaboration with several people, since then different parts can be
% edited independently.  This type of file organization is shown here.
%
\clearpage

\renewcommand{\thefootnote}{\arabic{footnote}}
\setcounter{footnote}{0}

%%%%%%%%%%%%%%%%
%\section{Introduction}
\label{sec:Introduction}
Lifetimes of $b$-flavored hadrons show the effects of all processes governing their weak decays.
In the case of neutral mesons, the decay rates are not purely exponential, but are modified by flavor mixing and  charge parity (\CP) violation.
The \Bsb meson's decay width $\Gamma_s$ differs for the heavy and light mass eigenstates, by an amount $\Delta\Gamma_s$ that has been measured to be significantly different from zero~\cite{PDG}.
This gives rise to a rich phenomenology of mixing and \CP violation. Precision measurement of the lifetime $\tau_s=\hbar/\Gamma_s$ is therefore an important benchmark.
The ratio of \Bsb to \Bzb lifetimes is well predicted in the heavy quark expansion (HQE) model~\cite{Lenz:2014jha}, which is used to extract values of the quark-mixing parameters $|V_{cb}|$ and $|V_{ub}|$, and thus lifetime measurements provide a precision test of the theory.

In this Letter we measure the lifetime of the decay $\brabar{B}_s^0\to D_s^{\pm}\pi^{\mp}$ by summing over \Bs and \Bsb states. 
Since \CP violation in \Bsb mixing is negligible \cite{Aaij:2013gta}, the final state receives equal contributions from light and heavy mass eigenstates.  Consequently, the decay rate
% Since this decay is characterized by equal amounts of light and heavy mass eigenstates, the decay rate
is given by the sum of two exponentials and can be fitted by a single exponential with the measured flavor-specific lifetime $\tau_{\rm fs}$ related to the decay width. 
Expanding in terms of $\Delta\Gamma_s/\Gamma_s$~\cite{Hartkorn:1999ga},\footnote{We use natural units where $\hbar=c=1$.}
\begin{equation}
\label{eq:DGcorr}
\tau_{\rm fs}\approx \frac{1}{\Gamma_s}\frac{1+\left(\frac{\Delta\Gamma_s}{2\Gamma_s}\right)^2}{1-\left(\frac{\Delta\Gamma_s}{2\Gamma_s}\right)^2}~.
\end{equation}
The \Bsb time-dependent decay rate is measured with respect to the well-measured  lifetimes of the \Bu and \Bdb mesons, which are reconstructed in final states
with similar topology and kinematic properties.\footnote{Reference to a given decay mode implies the use of the charge-conjugate mode as well.}

%%%%%%%%%%%%%%%%

%%%%%%%%%%%%%%%%
%\section{The LHCb detector}
\label{sec:Detector}
The \lhcb detector~\cite{Alves:2008zz} is a single-arm forward
spectrometer covering the \mbox{pseudorapidity} range $2<\eta <5$,
designed for the study of particles containing \bquark or \cquark
quarks. The detector includes a high-precision tracking system
consisting of a silicon-strip vertex detector surrounding the $pp$
interaction region~\cite{LHCb-DP-2014-001}, a large-area silicon-strip detector located
upstream of a dipole magnet with a bending power of about
$4{\rm\,Tm}$, and three stations of silicon-strip detectors and straw
drift tubes~\cite{LHCb-DP-2013-003} placed downstream of the magnet.
The tracking system provides a measurement of momentum, \ptot,  with
a relative uncertainty that varies from 0.4\% at low momentum to 0.6\% at 100\,GeV.
The minimum distance of a track to a primary vertex, the impact parameter, is measured with a resolution of $(15+29/\pt)\mum$,
where \pt is the component of \ptot transverse to the beam, in GeV.
Different types of charged hadrons are distinguished using information
from two ring-imaging Cherenkov detectors~\cite{LHCb-DP-2012-003}. Photon, electron and
hadron candidates are identified by a calorimeter system consisting of
scintillating-pad and preshower detectors, an electromagnetic
calorimeter and a hadronic calorimeter. Muons are identified by a
system composed of alternating layers of iron and multiwire
proportional chambers~\cite{LHCb-DP-2012-002}.

The trigger consists of a hardware stage, based on information from the calorimeter and muon systems, followed by a software stage, which applies a full event
reconstruction. The signal candidates are hardware triggered if there is at least one track having a large transverse energy deposit, then the track is required in software to have  a transverse momentum $\pt > 1.7$\,GeV and an impact parameter  $\chi_{\rm IP}^2$  with respect to the primary vertex (PV) greater than 16, where $\chi_{\rm IP}^2$ is defined
as the difference in $\chi^2$ of a given PV reconstructed with and without the considered particle included. In addition a vertex detached from the PV must be formed with either two, three or four tracks, with a scalar \pt sum of the tracks that must exceed a threshold that varies with the track multiplicity. 

  The advantage of measuring the \Bsb lifetime using the ratio with respect to well-measured lifetimes is that the decay time acceptances introduced by the trigger and selection almost cancel,
and only small corrections are required to the ratio of the decay time acceptances, which are taken from simulation. Thus, 
we reconstruct signals not only in the 
${\Bsb}\rightarrow{D^+_s \pi^-}$, $D^+_s \rightarrow K^+K^-\pi^+$ (denoted $\Bsb$$_{[KK\pi]}$) decay mode, but also in the topologically similar channels ($i$)
${B^-}\rightarrow{\D^0\pi^-}$, $\D^0\rightarrow K^-\pi^+$ ($\B^-_{[K\pi]}$), ($ii$)
${B^-}\rightarrow{\D^0\pi^-}$, $\D^0 \rightarrow K^-\pi^-\pi^+\pi^+$ ($\B^-_{[K\pi\pi\pi]}$),  and ($iii$)
${\Bzb}\rightarrow{D^+\pi^-}$, $D^+ \rightarrow K^-\pi^+\pi^+$ ($\Bzb_{[K\pi\pi]}$).  

%%%%%%%%%%%%%%%%
%\section{Event selection and mass fits}
\label{sec:Eventselection}

These decay modes are selected using some common criteria.
All of the tracks coming from candidate $D$ meson decays are required to have $\chi^2_{\rm IP} >9$. 
Pions arising from $\overline{B}$ meson decays have a more selective requirement $\chi^2_{\rm IP} > 36$ and they are required to be inconsistent with being identified as muons.
The $D$ candidates are required to have masses within 25\,MeV of their known values~\cite{PDG}, be reconstructed downstream of the PV,  and have $\chi^2_{\rm IP} > 4$.
The $D$ vertex separation from the $\overline{B}$ vertex should satisfy $\chi^2_{\rm VS} > 2$, where $\chi^2_{\rm VS} $ is the increase in $\chi^2$ of the parent $\overline{B}$ vertex
fit when the $D$ decay products are constrained to come from the $\overline{B}$ vertex, relative to when they are allowed to come from a separate vertex.

\Bm and \Bzb candidates are required to have $\chi^2_{\rm IP} <16$ with respect to the PV and masses in the ranges $5100-5600$\,MeV, while for \Bsb candidates the mass range is changed to $5200-5700$\,MeV.
The cosine of the angle between the $\overline{B}$ momentum and its direction of flight is required to be greater than 0.9999.
All signal candidates are refitted taking both $D$ mass and vertex constraints into account~\cite{Hulsbergen:2005pu}.
All charged particles are required to be identified as either pions or kaons.
Efficiencies are evaluated with a data-driven method using large samples of $\D^0 \rightarrow K^-\pi^+$ events, where the kinematic distributions of kaons and pions from the calibration sample are reweighted to match those of the $\overline{B}$ decays under study. 

We eliminate $\Bzb_{[K\pi\pi]}$ decay candidates that result from other similar decays, the ${\Bsb}\rightarrow{D^+_s \pi^-}$, $D^+_s \rightarrow K^+K^-\pi^+$  and ${\Lb}\rightarrow{\Lc \pi^-}$, $\Lc \rightarrow p K^-\pi^+$ modes, 
 if the invariant mass of the particles forming the $D^+$ candidate, with appropriately swapped mass assignments, is compatible within 30\,MeV with either of the known \Ds or \Lc  masses.
Similar vetoes are applied for $\Bsb$$_{[KK\pi]}$ candidates, where cross-feed from ${\Bzb}\rightarrow{D^+ \pi^-}$, $D^+ \rightarrow K^-\pi^+\pi^+$, and ${\Lb}\rightarrow{\Lc\pi^-}$, $\Lc \rightarrow p K^-\pi^+$
can happen if misidentification occurs. The combined efficiencies of the particle identification requirements and the mass vetoes depend on the specific decay mode considered, ranging from 80\% to 90\%, while more than 95\% of cross-feed backgrounds are rejected.

\begin{figure}[tb]
\begin{center}
\includegraphics[width=1.0\textwidth]{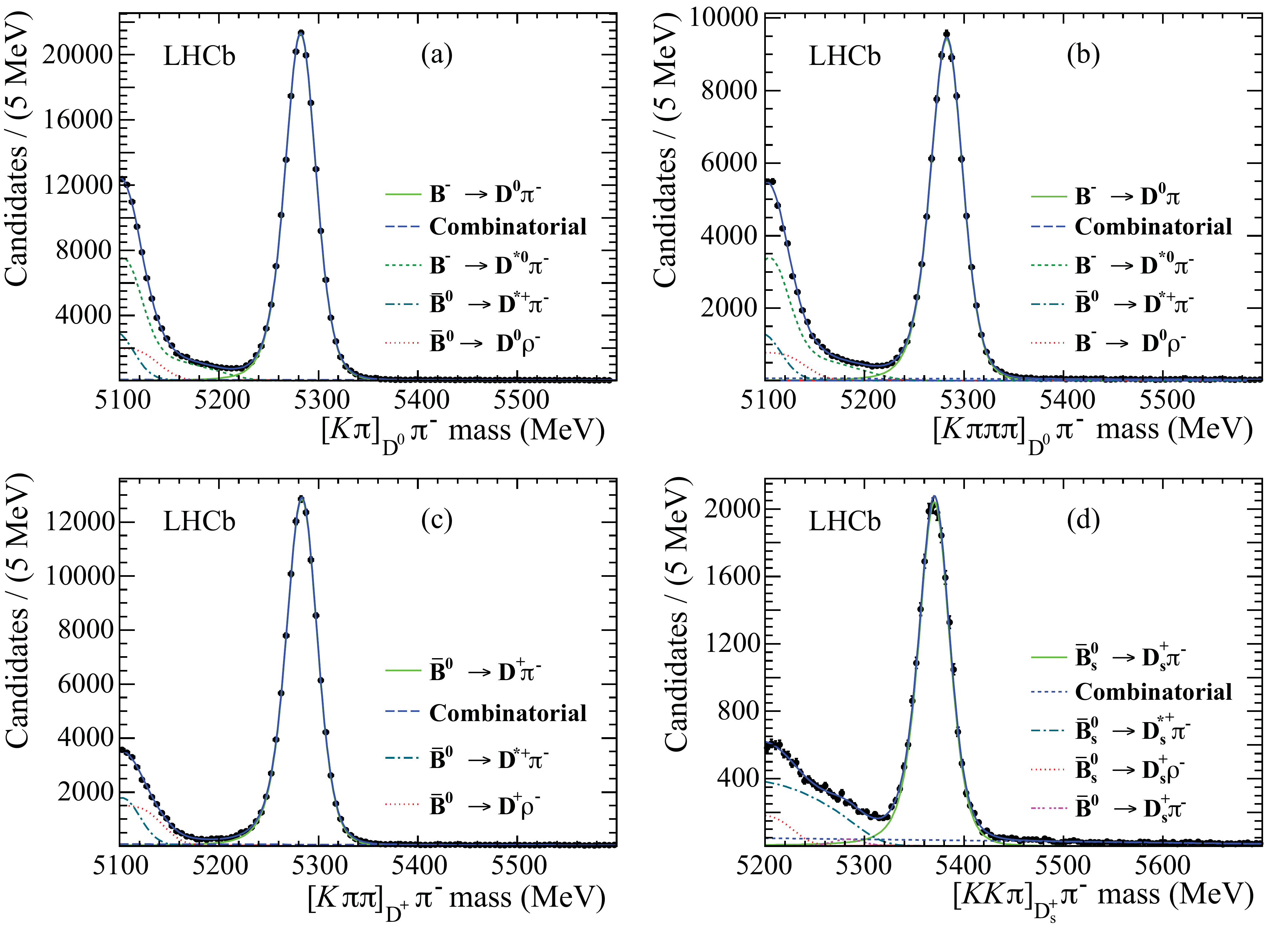}
\end{center}\label{fig1}
\vspace{-4mm}
\caption{\small Fits to the invariant mass spectra of candidates for the decays (a) ${B^-}\rightarrow{D^0[K\pi]\pi^-}$, (b) ${B^-}\rightarrow{D^0[K\pi\pi\pi]\pi^-}$, (c) ${\Bzb}\rightarrow{D^+[K\pi\pi]\pi^-}$, (d) ${\Bsb}\rightarrow{D^+_s[KK\pi]\pi^-}$.
The points are the data and the superimposed  curves  show the fit components. The solid (blue) curve gives the total. The $DK^-$ component is not visible, but is included.}
\end{figure}

The $\overline{B}$ candidate mass distributions for the four decay modes considered are shown in Fig.~\ref{fig1}, along with the results of binned maximum likelihood fits.
Signal shapes are parameterized using modified Gaussian functions (Cruijff) with independent tail shapes on both sides \cite{LHCb:2012fb,*delAmoSanchez:2010ae}.
All signal parameters are allowed to vary in the fit.
A residual component of $\overline{B} \rightarrow DK^-$ misidentified events is also included, with its yield constrained to that determined by an independent analysis where the kaon is positively identified.
Partially reconstructed backgrounds, where a pion or a photon are missed in reconstruction, are modeled using a sum of parametric empirical functions convolved with resolution functions. The unique decay kinematics of each of the modes, mostly determined by the polarization amplitudes, is taken into account. The combinatorial background is parameterized by a linear term. 
The fitted  signal yields are 
$179\,623 \pm 467$,
$82\,880  \pm 339$,
$109\, 670  \pm 378$ and
$21\,058  \pm 245$
 for 
$B^-_{[K\pi]}$,
$B^-_{[K\pi\pi\pi]}$,
$\Bzb_{[K\pi\pi]}$ and
$\Bsb$$_{[KK\pi]}$ decays, respectively.
%%%%%%%%%%%%%%%%
%%%%%%%%%%%%%%
%\section{Decay time acceptance}
\label{sec:time-res-accept}
The decay time, $t$, is derived from a flight-length measurement between production and decay points of the $\overline{B}$ particle, given by
\begin{equation}
t = m \cdot \frac{\vec{d} \cdot \vec{p}}{|p|^2} \, \mathrm{,}
\end{equation}
where $m$ is the reconstructed invariant mass, $\vec{p}$ the momentum and $\vec{d}$ the distance vector of the particle from its production to decay vertices. Prior to this determination, the PV position is refitted excluding the tracks forming the signal candidate, and the $\overline{B}$ meson is further constrained to come from the PV.
The decay time distribution of the signal $D_{\rm T}(t)$ can be described by an exponential function convolved with a decay time resolution function, $G(t,\sigma)$, and multiplied by an acceptance function $A(t)$:
\begin{equation}
D_{\rm T}(t) = A(t) \times [e^{-t'/\tau} \otimes G(t-t',\sigma)]. 
\end{equation}
The ratio of the measured decay time distributions of \Bsb to \Bdb or \Bm (we denote the use of either \Bdb or \Bu modes by the symbol $B_x$) can be written as
\begin{equation}
R(t) 
  = \frac{A_{\Bsb}(t) \times [e^{-t'/\tau_{\Bsb}} \otimes G(t-t',\sigma_{\Bsb})]}
             {A_{B_x}(t) \times [e^{-t'/\tau_{B_x}} \otimes G(t-t',\sigma_{B_x})]}.
\end{equation}
Resolutions are evaluated using simulated events and they are found to be 38, 37, 39 and 36\,\fs for $\Bsb$$_{[KK\pi]}$, $\Bzb_{[K\pi\pi]}$, $\B^-_{[K\pi]}$ and $\B^-_{[K\pi\pi\pi]}$, respectively.
Since the resolution is very similar in all the modes, and much smaller than our 0.5\,ps bin width, 
the resolution effects cancel \cite{Aaij:2014zyy}, and we are left with a ratio of two exponentials times the ratio of acceptance functions,
\begin{equation}
R(t) =  \frac{A_{\Bsb}(t)}{A_{B_x}(t)} e^{-t(1/\tau_{\Bsb}- 1/\tau_{B_x}) }= \frac{A_{\Bsb}(t)} {A_{B_x}(t)} e^{-t\Delta_{\Bs B_x}}, 
\end{equation}
where $\Delta_{\Bs B_x} \equiv 1/\tau_{\Bsb} - 1/\tau_{B_x}$.
Acceptance functions are evaluated by simulation.
The effective lifetime, $\tau_{\Bsb}$, can then be calculated from $\Delta_{\Bs B_x}$ using the well-known $B_x$ lifetimes. The current world average values are $\tau_{\Bzb}=1.519\pm 0.007$\,ps and $\tau_{\Bm}=1.641\pm 0.008$\,ps \cite{PDG}.

\begin{figure}[tb]
\begin{center}
%\vspace{-4cm}
\includegraphics[width=0.7\textwidth]{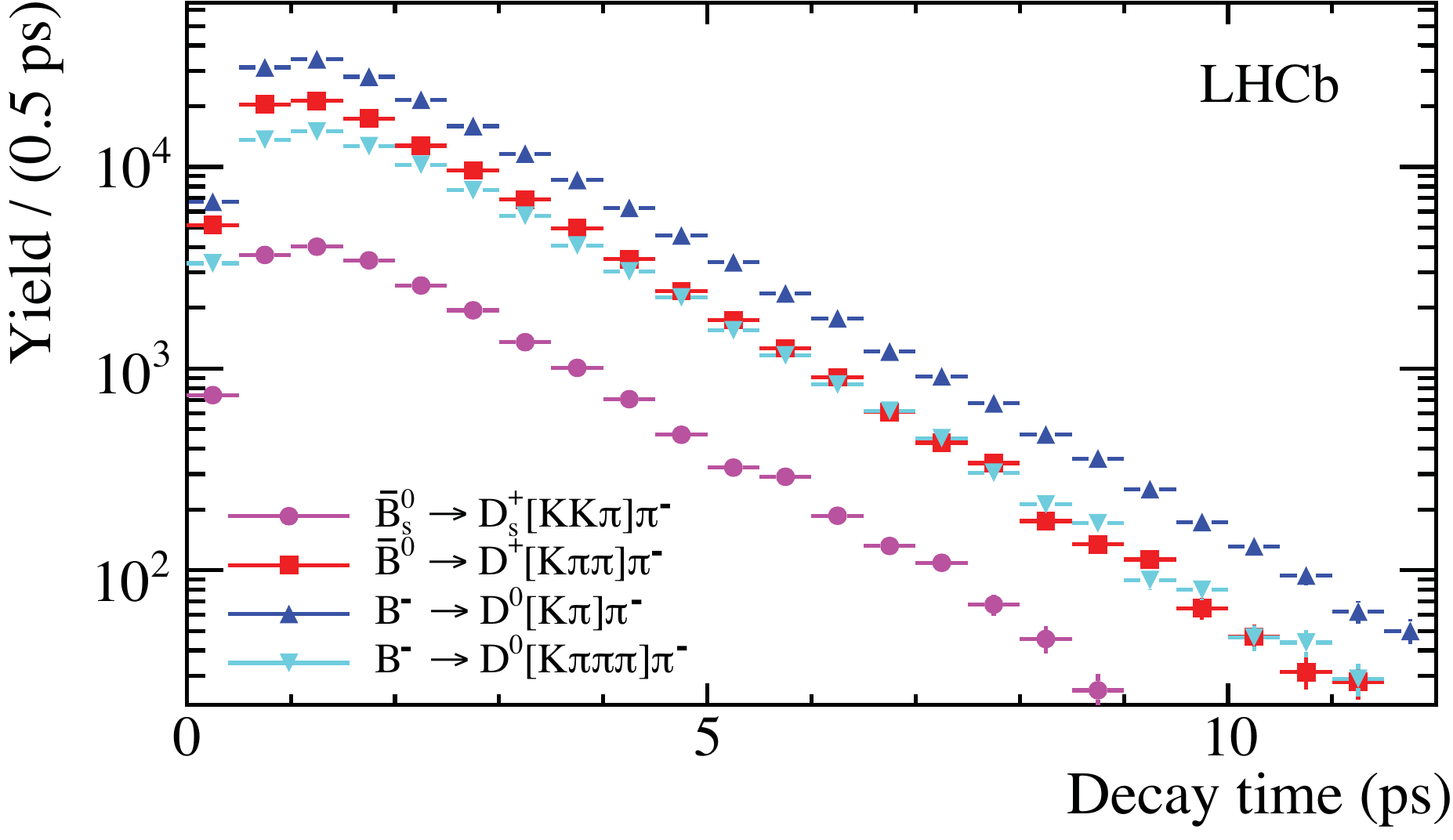}
\end{center}\label{fig2}
\vspace{-4mm}
\caption{\small Decay time distributions for ${B^-}\rightarrow{D^0[K\pi]\pi^-}$ shown as triangles  (blue), ${B^-}\rightarrow{D^0[K\pi\pi\pi]\pi^-}$  shown as inverted triangles (cyan), ${B^0}\rightarrow{D^+[K\pi\pi]\pi^-}$ shown as squares (red), ${\Bsb}\rightarrow{D^+_s[KK\pi]\pi^-}$ shown as circles (magenta).
For most entries the error bars are smaller than the point markers.}
\end{figure}

The signal yields are determined in each decay time bin by fitting the 
mass distribution in each bin with the same shapes as used in the full fits, with the 
signal shape parameters  fixed to those of the full fit as they are independent of the decay time.
The yields are shown in Fig.~\ref{fig2}.
%%

%$R(\bar{t}_i) = N_{\Bs}(\bar{t}_i) / N_{\Bx}(\bar{t}_i)$

\begin{figure}[tb]
\begin{center}
\includegraphics[width=0.7\textwidth]{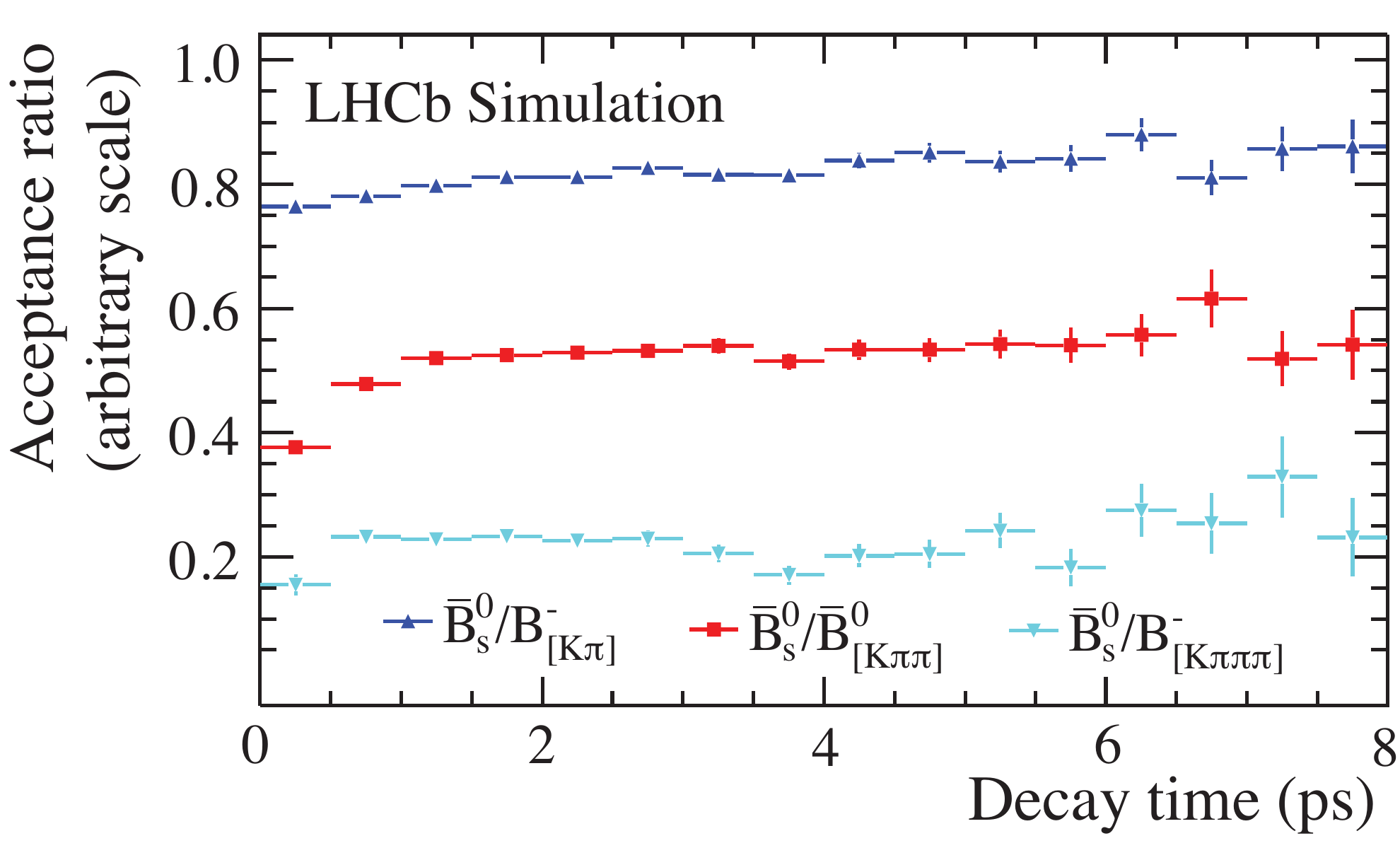}
\end{center}\label{fig3}
\vspace{-4mm}
\caption{\small Ratio of the decay time acceptances between ${\Bsb}\rightarrow{D^+_s[KK\pi]\pi^-}$ and ${B^-}\rightarrow{D^0[K\pi]\pi^-}$ shown as triangles (blue), ${B^0}\rightarrow{D^+[K\pi\pi]\pi^-}$ shown as squares (red), and ${B^-}\rightarrow{D^0[K\pi\pi\pi]\pi^-}$ shown as inverted triangles (cyan).
The vertical axis is shown in an arbitrary scale, different for each mode ratio to improve clarity.}
\end{figure}

The signal yields are then corrected by the relative decay time acceptance ratio, obtained by simulation, and shown in Fig.~\ref{fig3}. Then
the efficiency-corrected yield ratios are fitted with a single exponential function to extract $\Delta_{\Bs B_x}$. 
Fits are performed in the 1--8\ps region. The 0--1\ps region is excluded since the ratio of acceptances varies significantly here, due to the differences between the lifetimes and track multiplicities in the $D$ decays. 

%%%%%%%%%%%%%%%%

\begin{figure}[tb]
\begin{center}
\includegraphics[width=0.6\textwidth]{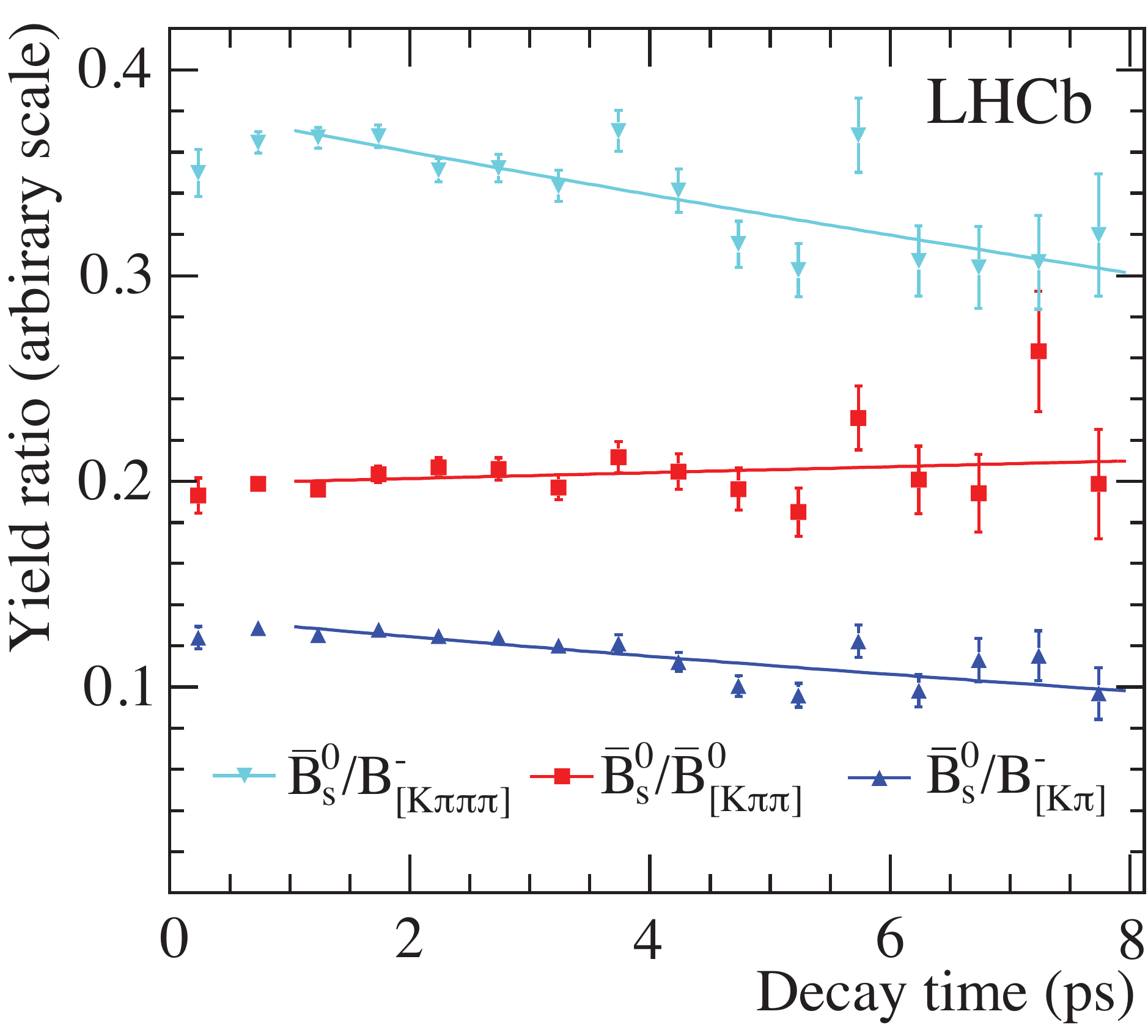}
\end{center}\label{fig4}
\vspace{-4mm}
\caption{\small Efficiency-corrected yield ratios of ${\Bsb}\rightarrow{D^+_s[KK\pi]\pi^-}$ relative to ${B^-}\rightarrow{D^0[K\pi]\pi^-}$ shown as triangles (blue), ${\Bdb}\rightarrow{D^+[K\pi\pi]\pi^-}$ shown as squares (red), and ${B^-}\rightarrow{D^0[K\pi\pi\pi]\pi^-}$ shown as inverted triangles (cyan). The simulation uncertainties are not included. 
The exponential fits are also shown. 
The vertical axis is shown in an arbitrary scale, different for each case to improve clarity.}
\end{figure}

%%%%%%%%%%%%%%%%
%\section{Results}
\label{sec:Results}
The full analysis is also applied to the control decay modes and the \Bu lifetime is measured relative to that of the \Bdb meson.
Given their well-known lifetimes, this provides a robust check on the validity of the procedure. We then measure the $\Bsb/B_y$ lifetime ratio for each of the three samples. The exponential fits for the $\Bsb/B_y$ lifetime ratios are shown in Fig.~\ref{fig4}, with
the results given in Table~\ref{tab:results}.
%%%
\begin{table}[tb]
\begin{center}
\caption{Measured lifetime ratios, compared with the known values, and the difference (fitted -- known), as well as the resulting measured lifetime $\tau_{\rm meas}$. Errors are statistical only. $B_x$ and $B_y$ indicate the modes used.}
\vskip 3mm
\label{tab:results}
%\begin{footnotesize}
\def\arraystretch{1.1}
\begin{tabular}{lccc}
\hline
Value                           &$\B^-_{[K\pi]}$/$\Bzb_{[K\pi\pi]}$ & $\B^-_{[K\pi\pi\pi]}$/$\Bzb_{[K\pi\pi]}$ & $\B^-_{[K\pi\pi\pi]}$/$\B^-_{[K\pi]}$ \\
\hline
Measured $\Delta_{B_x B_y}$ (\!\invps)           &$-0.0451\pm0.0033$  &$-0.0452\pm0.0039$  & $0.0011\pm0.0034$       \\
Known  $\Delta_{B_x B_y}$ (\!\invps)~\cite{PDG}&$-0.0489\pm0.0042$  &$-0.0489\pm0.0042$  & 0                        \\
Difference (\!\invps)                         &\,\,\,\,$0.0038\pm0.0054$   &\,\,\,\,$0.0037\pm0.0057$  &$0.0011\pm0.0034$ \\
$\tau_{\rm meas}(\Bm)$ (ps)     &\,\,\,\,$1.631\pm0.009$     & \,\,\,\,$1.631\pm0.010$   &   $1.638\pm0.009$ \\
\hline
Value                              & $\Bsb$$_{[KK\pi]}$/$\B^-_{[K\pi]}$ & $\Bsb$$_{[KK\pi]}$/$\Bzb_{[K\pi\pi]}$ & $\Bsb$$_{[KK\pi]}$/$\B^-_{[K\pi\pi\pi]}$ \\                         
\hline
Fitted $\Delta_{\Bs B_y}$           & $0.0402\pm0.0062$ & $-0.0063\pm0.0065$ & $0.0418\pm0.0066$ \\
$\tau_{B_y}$ (ps)~\cite{PDG}& $1.641\pm0.008$ & ~~$1.519\pm0.007$ & $1.641\pm0.008$\\
$\tau_{\rm meas}(\Bsb)$  (ps)           & $1.540\pm0.015$ & ~~$1.535\pm0.015$ & $1.535\pm0.016$ \\
\hline
\end{tabular}
%\end{footnotesize}
\def\arraystretch{1.0}
\end{center}
\end{table}
%%%
In each case good agreement with the known values of the light $\overline{B}$ meson lifetime ratio is found, and the three values of the \Bsb lifetime are consistent.

%\section{Systematic uncertainties}
\label{sec:sys}
The sources of systematic uncertainties on $\Delta_{\Bs B_x}$ are summarized in Table~\ref{t:SYSTEMATICS}.
The statistical precision on the relative acceptance is the largest source of systematic uncertainty.
The uncertainties due to the background description are estimated by comparing the nominal result to that obtained when the linear background slope is allowed to float separately in each decay time bin; in addition an exponential shape is used, and the largest
deviation is assigned as a systematic uncertainty.
Using a different signal shape to fit the data (double Crystal Ball function~\cite{Skwarnicki:1986xj}) leads to small changes.
There is also an uncertainty due to the decay time range and binning used. These uncertainties are ascertained by changing the fit range limits down to 0.5\ps and changing the size of the bins from 0.3\ps to 1\ps.
The relative measurements with respect to the three control samples agree within 0.005\ps,
and this is conservatively added to the total systematic uncertainty. 
\begin{table}[btp]
\begin{center}
%\begin{footnotesize}
\caption{Systematic uncertainties for $\Delta_{\Bs B_x}$. \label{t:SYSTEMATICS}}
\vskip 3mm
\def\arraystretch{1.1}
\begin{tabular}{lccc}
\hline
Source                            & $\Bsb$$_{[KK\pi]}$/$\B^-_{[K\pi]}$ & $\Bsb$$_{[KK\pi]}$/$\Bzb_{[K\pi\pi]}$ & $\Bsb$$_{[KK\pi]}$/$\B^-_{[K\pi\pi\pi]}$ \\ 
\hline
Lifetime acceptance               &0.003&0.004&0.005\\
Background model                  &0.002&0.002&0.002\\
Signal shape			  &0.0004&0.0005&0.0005\\
Binning schemes			  &0.003&0.001&0.005\\
\hline
Total				  &0.005&0.005&0.007\\
\hline
\end{tabular}
\def\arraystretch{1.0}
%\end{footnotesize}
\end{center}
\end{table}

%%%%%%%%%%%%%%%%
%\section{Conclusions}
\label{sec:Conclusions}
Using the known lifetimes of the \Bu and \Bdb mesons and the three different normalization channels, the flavor-specific \Bsb lifetime is determined as
\begin{equation}
\begin{array}{rcl}
\tau_{\rm fs}         &=& 1.540 \pm 0.015  \pm 0.012  \pm 0.008  \ps~[B^-_{[K\pi]}]\\
\tau_{\rm fs}     &=& 1.535 \pm 0.015  \pm 0.012  \pm 0.007  \ps~[\Bzb_{[K\pi\pi]}] \\
\tau_{\rm fs}  &=& 1.535 \pm 0.016  \pm 0.018  \pm 0.008  \ps~[B^-_{[K\pi\pi\pi]}], \\
\end{array}\nonumber
\label{e:dspiresult1}
\end{equation}
where the first uncertainty is statistical, the second systematic and the third the  uncertainty due to the input decay lifetimes of the \Bm and \Bdb mesons,  0.008\ps for the \Bm meson and 0.007\ps for the \Bd ~\cite{PDG}.
As the results are fully correlated, that with the smallest uncertainty is chosen
\begin{equation}
\tau_{\rm fs} = 1.535 \pm 0.015 \pm 0.012 \pm 0.007 \ps.\nonumber
\label{e:dspiresult2}
\end{equation}
This is the most precise measurement to date and it is consistent with previously available flavor-specific measurements~\cite{Aaltonen:2011qsa, PDG}, and measurements of \Bsb lifetimes in \CP eigenstate modes \cite{Stone:2014pra}. The lifetime ratio ${\tau_{\rm fs}(\Bsb)}/{\tau(\Bzb)}={1}/\left({1+\tau(\Bzb)\Delta_{\Bs\Bz}}\right)$ is determined as $1.010\pm0.010\pm0.008$, where we assign the uncertainty due to the \Bzb lifetime as purely systematic. A rather precise prediction of $\Gamma_d/\Gamma_s$ is given using the HQE model~\cite{Lenz:2014jha}. To compare with our measured lifetime ratio we apply
a 0.8\% correction from Eq.~(\ref{eq:DGcorr}) resulting in a corrected prediction for our measured lifetime ratio of $1.009\pm0.004$, in excellent agreement with our measurement, lending credence to this model.

We express our gratitude to our colleagues in the CERN
accelerator departments for the excellent performance of the LHC. We
thank the technical and administrative staff at the LHCb
institutes. We acknowledge support from CERN and from the national
agencies: CAPES, CNPq, FAPERJ and FINEP (Brazil); NSFC (China);
CNRS/IN2P3 (France); BMBF, DFG, HGF and MPG (Germany); SFI (Ireland); INFN (Italy);
FOM and NWO (The Netherlands); MNiSW and NCN (Poland); MEN/IFA (Romania);
MinES and FANO (Russia); MinECo (Spain); SNSF and SER (Switzerland);
NASU (Ukraine); STFC (United Kingdom); NSF (USA).
The Tier1 computing centres are supported by IN2P3 (France), KIT and BMBF
(Germany), INFN (Italy), NWO and SURF (The Netherlands), PIC (Spain), GridPP
(United Kingdom).
We are indebted to the communities behind the multiple open
source software packages on which we depend. We are also thankful for the
computing resources and the access to software R\&D tools provided by Yandex LLC (Russia).
Individual groups or members have received support from
EPLANET, Marie Sk\l{}odowska-Curie Actions and ERC (European Union),
Conseil g\'{e}n\'{e}ral de Haute-Savoie, Labex ENIGMASS and OCEVU,
R\'{e}gion Auvergne (France), RFBR (Russia), XuntaGal and GENCAT (Spain), Royal Society and Royal
Commission for the Exhibition of 1851 (United Kingdom).%%%%%%%%%%%%%%%%
%\bibliographystyle{lhcb}
%\bibliography{Dspi_onlypaper}
\input{Dspi.bbl}

% Author List ----------------------------
%  You need to get a new author list!
\newpage
\input{LHCb_HD_authorlist_2014-06-11.tex}
\end{document}

%% file: Dspi.bbl
\ifx\mcitethebibliography\mciteundefinedmacro
\PackageError{LHCb.bst}{mciteplus.sty has not been loaded}
{This bibstyle requires the use of the mciteplus package.}\fi
\providecommand{\href}[2]{#2}

%% file: LHCb_HD_authorlist_2014-06-11.tex
%%%%%%%%%%%%%%%%%%%%%%%%%%%%%%%%%%%%%%%%%%
\centerline{\large\bf LHCb collaboration}
\begin{flushleft}
\small
R.~Aaij$^{41}$, 
B.~Adeva$^{37}$, 
M.~Adinolfi$^{46}$, 
A.~Affolder$^{52}$, 
Z.~Ajaltouni$^{5}$, 
S.~Akar$^{6}$, 
J.~Albrecht$^{9}$, 
F.~Alessio$^{38}$, 
M.~Alexander$^{51}$, 
S.~Ali$^{41}$
% ...

{\footnotesize \it
% ...
$ ^{5}$Clermont Universit\'{e}, Universit\'{e} Blaise Pascal, CNRS/IN2P3, LPC, Clermont-Ferrand, France\\
$ ^{6}$CPPM, Aix-Marseille Universit\'{e}, CNRS/IN2P3, Marseille, France\\
% ...
$ ^{9}$Fakult\"{a}t Physik, Technische Universit\"{a}t Dortmund, Dortmund, Germany\\
% ...
$ ^{37}$Universidad de Santiago de Compostela, Santiago de Compostela, Spain\\
$ ^{38}$European Organization for Nuclear Research (CERN), Geneva, Switzerland\\
% ...
$ ^{41}$Nikhef National Institute for Subatomic Physics, Amsterdam, The Netherlands\\
% ...
$ ^{46}$H.H. Wills Physics Laboratory, University of Bristol, Bristol, United Kingdom\\
% ...
$ ^{51}$School of Physics and Astronomy, University of Glasgow, Glasgow, United Kingdom\\
$ ^{52}$Oliver Lodge Laboratory, University of Liverpool, Liverpool, United Kingdom\\
% ...
}
\end{flushleft}
%%%%%%%%%%%%%%%%%%%%%%%%%%%%%%%%%%%%%%%%%%